\documentclass[a4paper,11pt]{article}
\usepackage{pos}
\usepackage{xspace}

\usepackage[normalem]{ ulem }
\usepackage{soul}

\title{Limits on the Primordial Black Holes Dark Matter with current and future missions}
\ShortTitle{Current and future PBH limits}

\author*[a]{Denys Malyshev}
\author[b]{Emmanuel Moulin}
\author[a]{Andrea Santangelo}

\affiliation[a]{Institut f\"ur Astronomie und Astrophysik, Universit\"at T\"ubingen,\\
  Sand 1, D 72076 T\"ubingen, Germany}

\affiliation[b]{IRFU, CEA, D\'epartement de Physique des Particules, Universit{\'{e}} Paris-Saclay, \\
F-91191 Gif-sur-Yvette, France}

\emailAdd{denys.malyshev@astro.uni-tuebingen.de}
\emailAdd{emmanuel.moulin@cea.fr}
\emailAdd{andrea.santangelo@uni-tuebingen.de}

\abstract{In this proceeding we consider primordial black holes (PBHs) as a dark matter candidate. We discuss the existing limits on the fraction $\fpbh$ of the dark matter constituting of PBHs as a function of PBHs mass. The discussed limits cover almost all possible mass range with the currently only open window in $3\cdot 10^{16}-10^{18}$~g in which the PBHs can make up to 100\% of the dark matter content of the universe. We present the estimates of the capabilities of the near-future instruments (\ep/WXT, \svom/MXT) and discuss the potential of next-generation missions(\ath, \ths, \extp) to probe this mass range. We discuss the targets most suitable for the PBH dark matter searches with these missions and the potential limiting factor of the systematics on the derived results.}

\FullConference{
Multifrequency Behaviour of High Energy Cosmic Sources XIV (MULTIF2023)\\
12-17 June 2023\\
Palermo, Italy\\}

\def\gr{$\gamma$-ray\xspace}
\def\spi{INTEGRAL/SPI\xspace}
\def\xmm{XMM-Newton\xspace}
\def\extp{eXTP\xspace}
\def\ths{THESEUS\xspace}
\def\ath{Athena\xspace}
\def\mpbh{M_{\rm pbh}}
\def\fpbh{f_{\rm pbh}}
\def\ep{Einstein probe\xspace}
\def\svom{SVOM\xspace}
\def\xgis{THESEUS/XGIS\xspace}

\begin{document}
\maketitle

\section{Introduction}
Despite a vast body of evidences suggesting that the large fraction of the matter of the universe consists of dark matter (DM) not much is known about this specie except of its total density ($\Omega_{DM}h^2 = 0.1200 \pm 0.0012$, \cite{planck18}). For the first time the DM was invoked in early 1920s as an additional (to visible matter) component needed to explain the distribution of the stars velocities in our Galaxy~\cite{kapteyn22}. Ten years later DM was suggested to explain the unexpectedly high local density~\cite{oort32} and large velocities of the galaxies in the Coma cluster~\cite{zwicky33}. Since then, the number of evidences for the existence of DM has steadily increased, however its nature and most of its properties are still unknown even after a century of initial ideas, since only gravitational interaction of the DM with ``normal'' baryonic matter has been observed up to date. 

It is often assumed that DM is composed of particles that are either included in the Standard Model (SM) of elementary particles or yet unknown, but predicted by extensions/modifications of the SM. Astrophysical/cosmological observations already for a long time ruled out the first possibility. The baryonic nature of DM would imply a very high number of baryons in our Universe, contradicting cosmological nucleosythesis scenario, which otherwise correctly describes the observed abundances of the elements. Other non-baryonic DM candidates, e.g. neutrinos, are excluded by the large-scale structure and dwarf spheroidal galaxies observations (see e.g.~\cite{bond80} and~\cite{bertone05} for a review). The decades of the dedicated searches of the DM composing of many possible SM extensions also did not lead to any firm detection, see e.g.~\cite{pdg23} and references therein for a review.

In this proceeding we consider an alternative to the particles dark matter candidate -- primordial black holes (PBHs). Formed in the early universe these objects can satisfy all ``good DM candidate'' requirements: \textit{(i)}: to be produced in the early Universe (to be able to explain cosmic microwave background observations); \textit{(ii)}: to survive cosmological times (to explain present-day DM); \textit{(iii)}: be cold, at most warm (to explain the structure formation); \textit{(iv)}: be massive enough to gravitationally interact with the baryonic matter. The interest to PBHs has been also significantly increased in the last years in light of a detection of the gravitational waves with from the merging black holes with LIGO/VIRGO experiment~\cite{LIGOScientific:2016aoc}. 

In what below we discuss the existing constraints on the fraction $\fpbh$ of the dark matter that could be made of PBHs derived with the currently operating instruments. We show that in a relatively narrow PBHs mass range $3\cdot 10^{16}- 10^{18}$~g these peculiar objects could make up to 100\% of the observed DM, not violating any DM-relevant constraints. We discuss the capabilities of the future X-ray missions (\ths, \extp, \ath) to probe this mass range. For the first time we discuss also the limits that could be achieved with the near-future missions such as \ep and \svom that are expected to see the first light at the end 2023 -- beginning 2024. We discuss the most suitable targets for observations with these missions and the impact of the systematic uncertainties connected to the mis-modeling of the instrumental/astrophysical background on the derived limits.

\section{Existing PBH constraints}
\begin{figure}
    \centering
    \includegraphics[width=\textwidth]{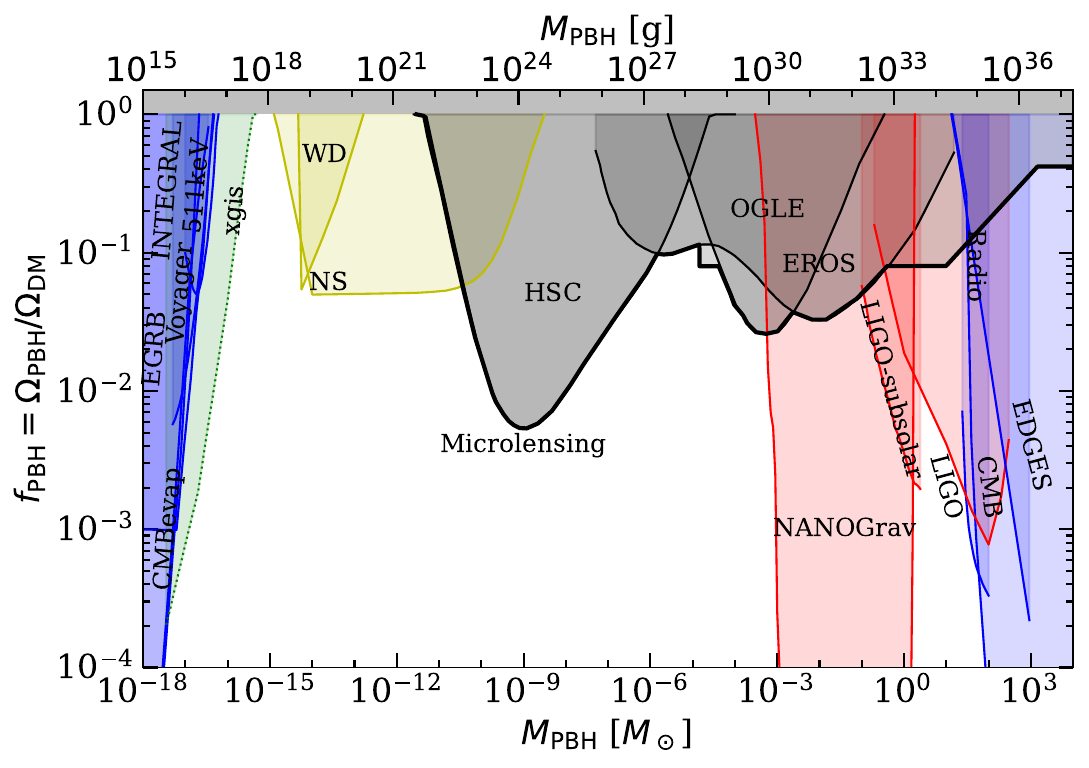}
    \caption{The existing constraints on the fraction of the dark matter $\fpbh$ consisting of the primordial black holes as a function of PBH mass $\mpbh$. Please note that the green dotted-line limited region shows the expected limits that could be derived with the future next-generation mission \ths. The Figure is generated with the help of \href{https://github.com/bradkav/PBHbounds}{PBHbounds} code~\cite{bradley_j_kavanagh_2019_3538999}.}
    \label{fig:all_constraints}
\end{figure}

In the simplest case the non-rotating primordial black holes constituting the dark matter could be characterised by only two parameters -- the mass of the PBH $\mpbh$ and the fraction of the dark matter made of these black holes $\fpbh$. In absence of a-priori knowledge of the exact $\mpbh$ value all existing constraints are shown for $\mpbh-\fpbh$ parameter space, see e.g. Fig.~\ref{fig:all_constraints} where the shaded areas correspond to the excluded $\fpbh$ values.

We would like to note also, that generally speaking PBHs could be characterised by a non-trivial mass and momenta distribution, see e.g.~\cite{Carr:2017jsz,Kuhnel:2017pwq,Arbey:2019vqx}. For the sake of simplicity we consider below only the case of monochromatic mass-function and non-rotating PBHs, noting that the not-trivial mass/momenta functions could relax the constraints on these objects by a factor of few~\cite{ray21}.

%%%%%%%%%%%%%%%%%%%%%%%%%%%%%%%%%%%%%%%%%%%%%%%%%%%%%%%%%%%%%%%%%%
%%%%%%%%%%%%%%%%%%%%%%%%%%%%%%%%%%%%%%%%%%%%%%%%%%%%%%%%%%%%%%%%%%

In what below we summarize and briefly discuss the constraints on PBHs derived by date by currently operating missions/experiments. These constraints can be broadly sub-divided over the several classes applicable for the following characteristic PBH mass range:
\begin{itemize}
    \item PBHs evaporation signatures-based constraints, ($\mpbh\lesssim 3\cdot 10^{16}$~g)
    \item NS/WD-detonation based constraints ($\mpbh \sim 10^{19}-10^{23}$~g)
    \item Microlensing-events based constraints ($\mpbh\gtrsim 10^{22}$~g)
    \item Gravitational-waves based constraints ($\mpbh\sim 10^{30}-10^{35}$~g)
    \item Early-universe constraints ($\mpbh\gtrsim 10^{35}$~g)
\end{itemize}
The current limits on $\fpbh$ as a function of $\mpbh$ for the broad range of masses are shown schematically in Fig.~\ref{fig:all_constraints}.
In what below we briefly describe each of these constraints.
%%%%%%%%%%%%%%%%%%%%%%%%%%%%%%%%%%%%%%%%%%%%
\paragraph{PBHs evaporation signatures-based constraints.} If exist in the present-day Universe PBHs are expected to evaporate producing Hawking radiation~\cite{Hawking:1974rv}. The evaporation PBH time
\begin{align}
& \tau_{\rm evap}\sim 10^{67}\left(\frac{\mpbh}{M_{\odot}}\right)^3\,\mbox{yr}    
\end{align}
for the PBHs with masses $\lesssim 10^{14}$~g is comparable or shorter than the lifetime of the Universe~\cite[see e.g.][]{villanueva21}. This does not allow one to suggest the very light PBH contributing significantly to the present day dark matter. The PBHs with slightly higher masses of $\mpbh\lesssim 10^{15}$~g are expected to evaporate in the present-day Universe producing the burst of the high-energy emission. The non-detection of such bursts in the TeV band puts strong limits on the $\fpbh$ for the light primordial black holes~\cite{hess_pbh23}. 

\begin{figure}
    \centering
    \includegraphics[width=\textwidth]{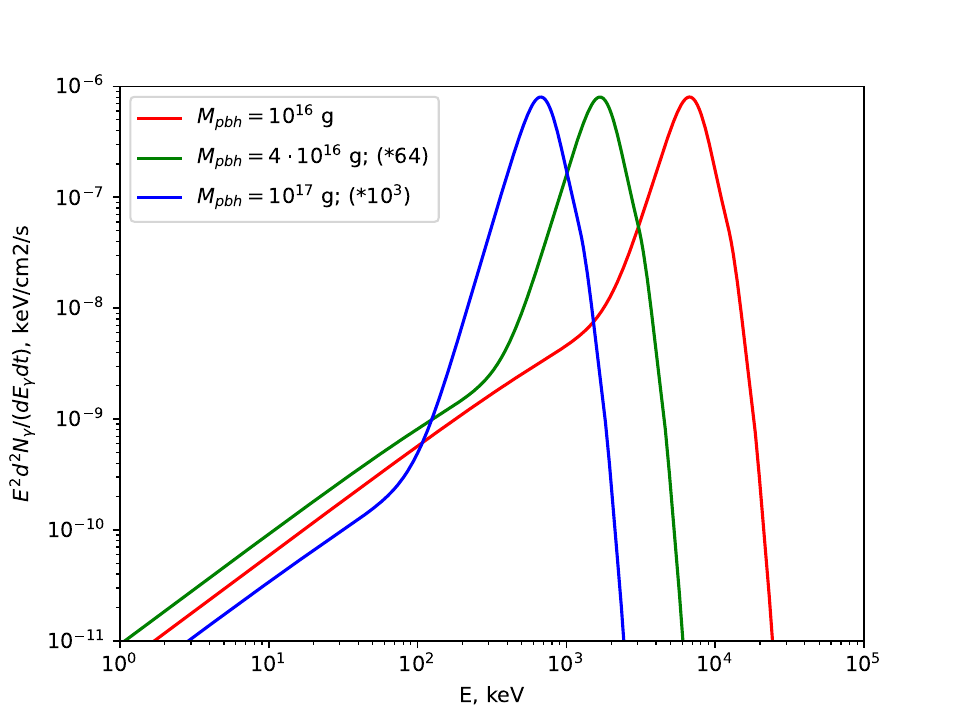}
    \caption{The expected signal ($E^2d^2N_\gamma/(dE_\gamma dt)$, see Eq.~\ref{eq:hawking_signal}) from the evaporating PBH for the several $\mpbh$ masses. Note, that the signals from the PBHs with $\mpbh=4\cdot 10^{16}$~g and $\mpbh=10^{17}$~g are rescaled (multiplied) by factors 64 and 1000 correspondingly.}
    \label{fig:model_signal}
\end{figure}

The PBHs with higher masses could survive the present-day Universe, but still producing steady in time Hawking radiation with the characteristic temperature
\begin{align}
& T_{\rm H} = 1/(4\pi/G_{N\rm }\mpbh)\simeq 1.06\times(10^{16} \rm g / \mpbh)\,\mbox{MeV}    
\label{eq:hawking_T}
\end{align}
is in the keV-MeV range (see e.g. Fig.~\ref{fig:model_signal}) for the PBHs with masses $\gtrsim 10^{16}$~g. 
The non-detection of this radiation from certain DM-dominated objects with the current X-ray/$\gamma$-ray instruments could be used to put constraints on $\fpbh$. The expected strength of the signal is proportional to the $\fpbh$ and the total amount of dark matter on the line of sight to this object. Please note also, that the strength of the signal drops as $\sim \mpbh^3$ (see e.g. Fig.~\ref{fig:model_signal}) which challenges the direct application of this method to the high-masses PBHs despite the fact that the maximum of the signal could remain in the keV band accessible for modern space missions.

The non-detection of a Hawking radiation from the dark matter dominated regions of the MW and Draco dSph with INTEGRAL/SPI and \xmm allowed~\cite{pbh_we} to constrain $\fpbh<0.1$ for $\mpbh<3\cdot 10^{16}$~g. The best expected limits from the next-decade missions (\ths limits) are shown in Fig.~\ref{fig:all_constraints} with the green dotted-line limited region~\cite{pbh_we}, see also Fig.~\ref{fig:ath_extp} and Fig.~\ref{fig:ep_svom_spi} for the limits expected from other future missions. %We would like to note also the important role of the systematic uncertainty of the instrumental background modelling which for~\cite{pbh_we} was the limiting factor for $\fpbh$ constraints.

Similarly, PBHs during their evaporation could produce the electron-positron pairs. The produced in such a way positrons could annihilate with the ambient-medium electrons which leads to the production of the $511$~keV \gr line. \cite{laha19} used \spi observations of such a line from the GC vicinity and derived the limits on the $\fpbh$ requiring that the observed flux in 511~keV line does not exceed the modelled one. 

The observations of the featureless, powerlaw extragalactic isotropic gamma-ray background (EGRB) in the keV-MeV band could also be used to constrain the fraction of DM made of PBHs~\cite{carr10,iguaz21}. Sufficiently high $\fpbh$ values for the PBHs constituting the majority of the DM in the MW
would result in observable distortions of the powerlaw spectral shape of the EGRB, as the line of sight for all EGRB measurements intersects the MW. Another constraints in $\mpbh\lesssim few\times 10^{16}$~g originate from the non-observation of CMB anisotropies damping and recombination history changes expected due to an additional heating from the Hawking radiation of the evaporating low-mass PBHs in the early Universe~\citep{clark17}.

Another low-PBH mass limit originates from the non-observation of the increase of the sub-GeV $e^\pm$ flux by Voyager~1~\citep{boudaud19}. As discussed above PBHs with the masses $\mpbh\lesssim 10^{16}$~g during their evaporation are expected to inject to the medium MeV energy-scale electrons/positrons. Such low-energy particles are effectively deflected inside of the Solar System by the solar magnetic field, which complicates their studies by the local to the Earth missions. However, $e^\pm$ fluxes beyond the shielded by the magnetic field region can be measured by Voyager~1 which has passed the heliopause in 2012. The non-observation of the $e^\pm$ flux increase allowed \cite{boudaud19} to put limits on the local to the Solar System amount of the primordial black holes.
%%%%%%%%%%%%%%%%%%%%%%%%%%%%%%%%%%%%%%%%%%%%%%%%%%%%%%%%%%%%%%%%%%%%%%%%%%
\paragraph{NS/WD-detonation based constraints.}
The constraints for higher-mass PBHs ($\mpbh\sim 10^{19}-10^{23}$~g) are based on the possible thermonuclear detonations or destroyment of the neutron stars(NS)/white dwarves(WD) as these objects are occasionally crossed by a PBH~\citep{capela13,graham15}. Recently the same approach was considered in case of the detonation of the Sun-like stars in dwarf spheroidal galaxies~\citep{esser23}. We would like to note, however, that these constraints are strongly model dependent and are actively debating in the literature, see e.g.~\cite{capela14,defillon14,camacho19}.
%%%%%%%%%%%%%%%%%%%%%%%%%%%%%%%%%%%%%%%%%%%%%%%%%%%%%%%%%%%%%%%%%%%%%%%%%
\paragraph{Microlensing events based constraints.}
The fraction of PBHs with masses higher than $\sim 10^{22}$~g is significantly constrained by non-observation of the large number of microlensing events. The PBH (or any gravitating body) passing between the observer and a bright distant object (e.g. a star) would cause the rapid increase of the brightness of this object due to microlensing effect, see e.g.~\cite{mao12} for a review. The amplitude of the flux magnification and the characteristic time scale of the microlensing event increase with the mass of the PBH. For the typical kpc-scale distances between the distant stars and PBHs, the characteristic duration of the magnification event is of order of 20 days for $\mpbh\sim 10^{33}$~g and scales as $\mpbh^{1/2}$ with the PBH mass. 
Typical targets for the search of microlensing events are nearby galaxies (e.g. M~31 or LMC) which allow the observation of a large number of distance stars simultaneously and perform corresponding searches of microlensing events in parallel. We would like to note, that the PBH mass range to which such surveys are sensitive is limited from below by short variability timescales which can be confused with the own variability of the stars. The sensitivity to highest mass PBHs is limited by a long variability timescale, exceeding the duration of observational projects. In addition the total number of expected microlensing events rapidly decreases with the increase of the $\mpbh$ as lower number of more massive PBHs is required to explain the constant amount of the DM in the distant object, see e.g.~\cite{niikura19}. 

The tightest constraints from the non-observations of the microlensing events originate from the Subaru/HSC optical survey of the M~31~\cite{croon20}; EROS observations of the Magellanic Clouds~\cite{eros}; 5-years long observations of the Galactic Bulge region with OGLE~\cite{ogle}, see Fig.~\ref{fig:all_constraints}.
%%%%%%%%%%%%%%%%%%%%%%%%%%%%%%%%%%%%%%%%%%%%%%%%%%%%%%%%%%%%%%%%%%%%%%%%%
\paragraph{Gravitational-waves based constraints} 
The PBHs with masses from a fraction to hundreds of solar masses can be constrained by the non-detection of the gravitational waves signal from these objects. In case of the relatively light PBHs with the masses of $10^{30}-10^{33}$~g such signal could be produced during their formation and detected with the Pulsar Array/NANOGRAV~\cite{nanograv20}. 
A fraction of PBHs with the subsolar masses of $10^{32}-10^{33}$~g are expected to be in a binary systems; the non-detection of such objects with LIGO allowed~\cite{ligo1} to put constraints on such objects. The objects with masses in the range of $10^{33}-10^{36}$~g, \textit{i.e.} $1-100 M_{\odot}$, were detected by LIGO in binary systems; however the significantly larger amount of such objects was argued to be detected in case if PBHs constitute the significant fraction of the DM~\cite{kavanagh18} which puts tight constraints on PBHs of such masses.
%%%%%%%%%%%%%%%%%%%%%%%%%%%%%%%%%%%%%%%%%%%%%%%%%%%%%%%%%%%%%%%%%%%%%%%%%
\paragraph{Early-Universe constraints}
The fraction of the dark matter constituting of the most massive PBHs with masses $\mpbh\gtrsim 10^{35}$~g is limited by their impact on the early-universe observables. The tight limits are based on the non-detection of the spectral~\cite{kohri14} and spatial~\cite{serpico20} distortions of the CMB resulted by an accretion on the primordial black holes. Yet another limits originate from the observations of a relatively cold hydrogen at $z\sim 17$ with EDGES~\cite{hektor18}; the presence of the PBHs would result in the accretion of the hydrogen with its consecutive heating.

\begin{table}[]
    \centering
    \resizebox{\linewidth}{!}{%
    \begin{tabular}{c|c|c|c|c|c|c|c}
    \hline\hline
    Instrument    & Energy range  & Peak $A_{\rm eff}$  & FoV & Launch date & Target & Obs. Type &$D$-factor\\
                  & [keV]           & [cm$^{2}$]        & [sr]  & [year]   &    &    & [GeV/cm$^2$]\\
                  \hline\hline
       \xmm/PN       &     0.1-15    & 815             & $4.5\cdot 10^{-5}$ & 1999-**& Draco+MW& Model  & $(1.1+0.74) \cdot 10^{18}$\\%$1\cdot 10^{18}$\\
    \spi          &     20-8000   & 160             & 0.29               & 2002-**& MW & ON-OFF     &$0.9\cdot 10^{22}$\\
    \hline
    \extp/SFA     &   0.5-10      & 8600            & $9.6\cdot 10^{-6}$ & 2027& Segue I + MW&Model& $(2.0+0.9) \cdot   10^{17}$\\%$2.84\cdot 10^{17}$ \\            
    \extp/LAD     &   2-30        & $3.3\cdot 10^4$ & $2.4\cdot 10^{-4}$ & 2027 &  Segue I& ON-OFF & $9.8\cdot 10^{17}$\\
    \extp/WFM     &     2-50      & 77              & 2.5                & 2027 &  MW & ON-OFF    & $2\cdot 10^{22}$ \\
    \ths/SXI      &     0.3-5     & 1.9             & 1                  & 2037& MW & ON-OFF     &$1\cdot 10^{22}$\\
    \ths/XGIS-X   &     2-30      & 504             & 1                  & 2037& MW & ON-OFF     &$1\cdot 10^{22}$\\
    \ths/XGIS-S   &     20-2000   & 1060            & 1                  & 2037& MW & ON-OFF     &$1\cdot 10^{22}$\\
    \ath/X-IFU    &     0.2-12    & $1.6\cdot 10^4$ & $3.3\cdot 10^{-6}$ & 2035& Segue I+MW &Model & $(8.3+3.0)\cdot 10^{16}$\\
    \ath/WFI      &     0.2-15    & 7930            & $1.35\cdot 10^{-4}$& 2035& Segue I+MW &Model &$(0.98+1.2)\cdot 10^{18}$\\ \hline
    \ep/WXT      &     0.5-4    & 3            & $1.1$& 2023& MW &ON-OFF &$1\cdot 10^{22}$\\
    \svom/MXT    &     0.2-10   & 37           & $3.5\cdot 10^{-4}$& 2023-24& Segue I &ON-OFF &$0.98\cdot 10^{18}$\\
    \hline
    \hline 
    \end{tabular}}
    \caption{The characteristics of the currently operating and future missions considered in this work. The table summarizes the operating energy range of the instrument, peak effective area, the field of view, and the planned launch date (as of August 2022). The last columns summarize the target, the type of (proposed) observation (background Model or ``ON-OFF''), and the estimated $D$-factor. For the ``ON-OFF''-type observations the $D$-factor corresponds to the $D$-factor difference between the ON ad OFF regions. Adopted D-factor values are based on~\cite{Geringer-Sameth:2014yza} and~\cite{Cautun:2019eaf}. The parameters of \xmm, INTEGRAL, \extp, THESEUS and \ath missions are adopted from~\cite{pbh_we}.
    }
    \label{tab:future_missions}
\end{table}

\subsection{Constraints with future instruments}
\begin{figure}
    \centering
    \includegraphics[width=0.47\textwidth]{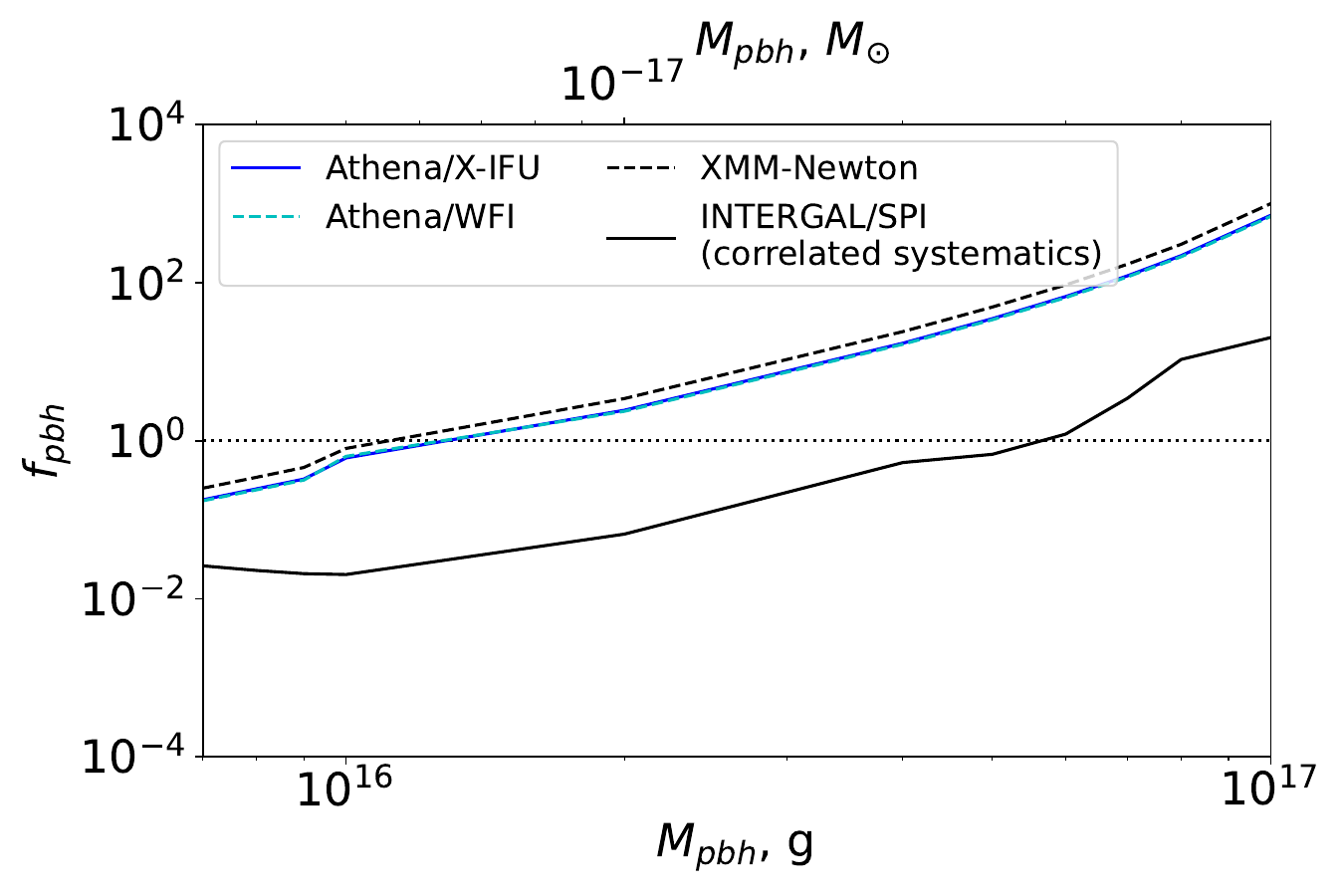}
    \includegraphics[width=0.47\textwidth]{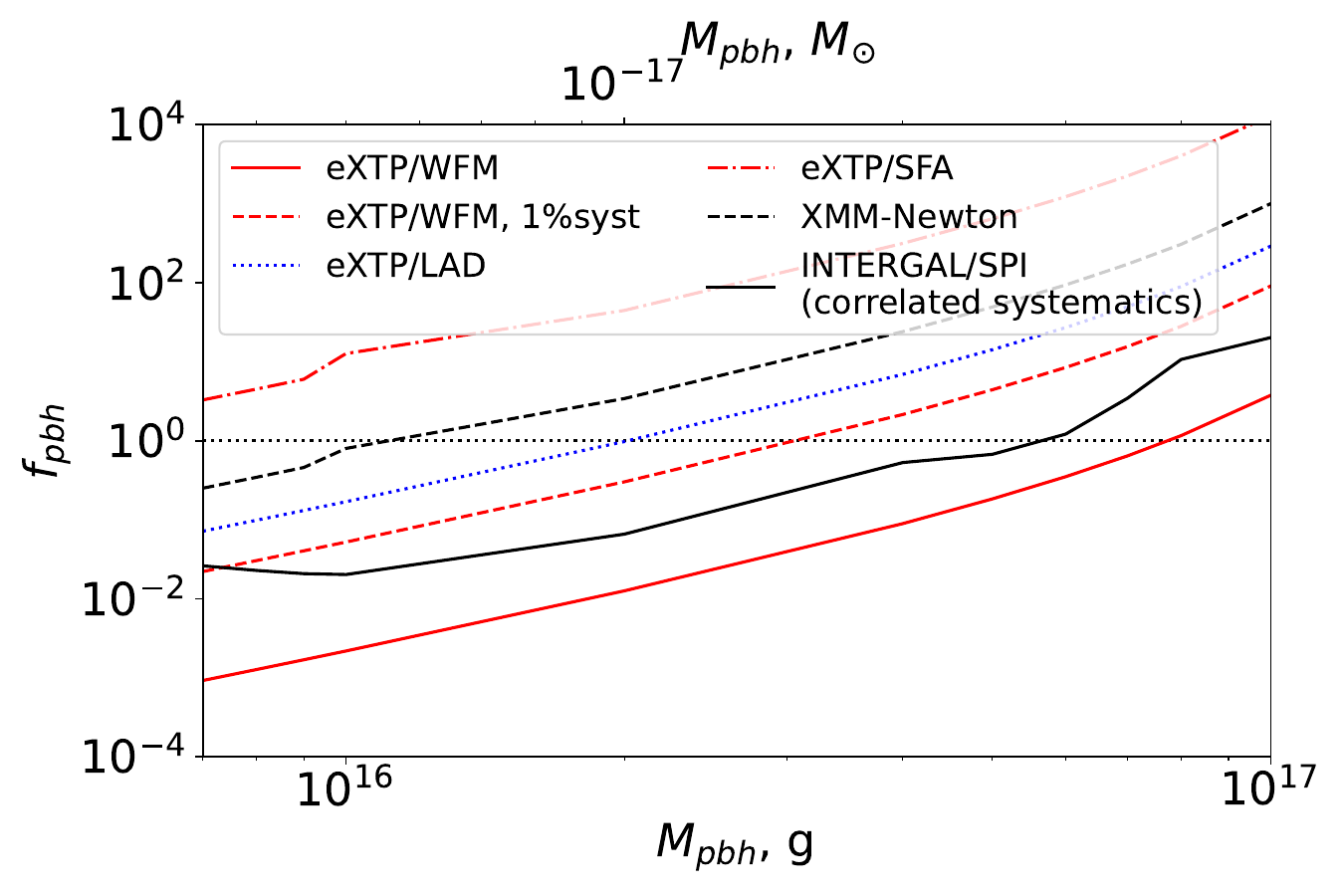}
    \caption{The expected limits on the fraction of DM consisting of evaporating PBHs based on \ath and \extp proposed observations of the Segue~I dSph and MW dark matter dominated regions correspondingly. The \xmm limits are based on the non-detection of the signal from evaporating PBHs in $\sim 1$~Msec long observations of Draco dSph. The black line corresponds to the limits derived with \spi observations of the close to the GC regions. The Figure is adopted from~\cite{pbh_we}. }
    \label{fig:ath_extp}
\end{figure}

The constraints discussed above relatively well cover the whole PBH mass range from $\lesssim 10^{15}$~g to $\gtrsim 10^{36}$~g. Currently the only mass range for which the fraction of the dark matter constituting of the PBHs is almost not constrained\footnote{Note, that the strong GRB femtolensing constraints present at the lower edge of this band were recently debated and significantly relaxed~\cite{Katz:2018zrn}} is limited to $\gtrsim 3\times 10^{16}-10^{18}$~g. At least partially this window could be closed with the observations of the (steady) Hawking radiation 
from the currently evaporating PBHs. %\textbf{Do you implicitly refer here to extended mass function for currently evaporateing PBHs ? } 
As discussed above for the PBHs of such masses the Hawking radiation is expected to be characterised by a keV-MeV temperature potentially detected by a present or next-generation X-ray missions. E.g.~\cite{pbh_we} shown that the non-detection of the Hawking radiation with INTEGRAL/SPI observations of the inner Galaxy regions as well as the observations of the dwarf spheroidal galaxies with \xmm allow to constrain the $f_{pbh}$ to be smaller then several per cent for the masses $\mpbh\lesssim 2\cdot 10^{16}$~g. The authors argued that this limit can be substantially improved by several next-generation missions, see Fig.~\ref{fig:ep_svom_spi} and Tab.~\ref{tab:future_missions} for the main characteristics of the considered missions. Many of these missions are characterised by a large field of view (FoV) and/or relatively high effective area in comparison to the currently operating ones. The observations of the MW regions with the low contributions of the convenient astrophysics sources with the large-FoV missions, in particular with \xgis could allow us to improve the existing limits  by up to two orders of magnitude (depending on the level of control of systematic uncertainties), see Fig.~\ref{fig:ep_svom_spi}. The main limiting factor for the derived with these missions limits was found the systematic uncertainty~\cite{pbh_we}. The poor modelling of the strong time-variable instrumental or a complex astrophysical background could result in under-estimation of the corresponding uncertainties and lead to artificially strong constraints. 
The narrow-FoV, large effective area present day (\xmm) and future missions (\ath and \extp) may not be very efficient for limiting the evaporating PBHs~\cite{pbh_we}, see Fig.~\ref{fig:ath_extp}.

\subsubsection{Future constraints with \ep and \svom}
Here we consider another two missions, expected to be launched in the near future (end 2023 -- beginning 2024) -- \ep and \svom and discuss their perspectives for the putting constrains on the fraction of DM that could constitute of PBHs -- $f_{pbh}$.

\paragraph{EP} (\ep) is a soft X-ray Chinese-European mission~\cite{einstein_probe} expected to be launched at the end of 2023. It is designed as a full-sky monitor and will host Wide-Field (WXT) and Follow-up (FXT) X-ray Telescopes. The WXT is characterised by a large FoV of $\sim 3600$~deg$^2$, an effective area of $\sim 3$~cm$^2$ and designed as a primary all-sky scanning instrument ($\sim 5'$ angular resolution) operating in $0.5-4$~keV band. The FXT is an \xmm-class telescope operating in 0.3--10~keV band, FoV of $\sim 1$~deg$^2$ and effective are of $\sim 600$~cm$^2$. The FXT angular resolution is expected to be $\sim 4''$~\cite{einstein_probe}. As the FXT instrument characteristics are close to those of \xmm and the \xmm capabilities for the PBH searches were discussed in \cite{pbh_we}, in what follows we focus on the WXT instrument of this mission.  

\paragraph{\svom}(Space-based multi-band astronomical Variable Objects Monitor) is a Chinese-French mission designed mainly for the detection, localization and follow-up observations of the Gamma-Ray Bursts and other high-energy transients~\cite{svom}. As of August 2023 the mission is planned to be launched at the end of 2023--beginning 2024. \svom will host onboard Gamma-Ray Monitor (GRM), hard X-ray instrument ECLAIR, a microchannel X-ray telescope (MXT) and a visual-band telescope (VT). 

The GRM instrument is expected to operate in 15~keV -- 5~MeV energy range and will consist of 3 modules looking at different direction on the sky and characterised by the effective areas of $\sim200$~cm$^2$ and field of view $\lesssim 1$~sr each. The primary science goal of this instrument is the detection of gamma-ray bursts (detection of 90 GRBs per year is expected~\cite{svom}). 

The ECLAIR instrument is operating in 4-150~keV energy range, effective area of $\sim 1000$~cm$^2$ and field of view of $\sim 2$~sr. The instrument is designed as a coded-mask hard X-ray imager and a trigger for the \svom mission.

The microchannel X-ray telescope (MXT) is a focusing soft-X-ray instrument. It is expected to operate in 0.2-10~keV energy range, have a FoV of $64'\times 64'$ and effective area of $\sim 37$~cm$^2$. The observations of the GRBs aftrglows with this instrument are expected to significantly improve the localisation accuracy of these events with ECLAIR. 

In what below we discuss the capabilities of \ep/WXT and \svom/MXT instruments in a context of their capabilities of putting limits on the evaporating PBHs in a way similar to the one discussed for the next-generation missions in~\cite{pbh_we}. The basic characteristics of these instruments are summarized and compared to the rest of the missions discussed in~\cite{pbh_we} in Tab.~\ref{tab:future_missions}. Following the approach proposed in~\cite{pbh_we} we propose to derive the constraints basing on searches of the Hawking radiation from the evaporating PBHs constituting dark matter in certain DM-dominated regions. 

\paragraph{The expected PBH-evaporation signal} (Hawking radiation) from a dark matter dominated object is characterised by a non-trivial spectrum\footnote{~Please note, that the signal is steady in time for $\mpbh\gtrsim 10^{15}$\,g. The lower masses PBHs are expected to evaporate in the present-day Universe producing a burst of high-energy emission, see e.g.~\cite{villanueva21, hess_pbh23}}~\cite{Hawking:1974rv}:
\begin{align}
\label{eq:hawking_signal}
& \frac{d^2\Phi_{\gamma}}{dE_{\gamma}}(\Delta \Omega) = \frac{1}{4\pi} \int\limits_{\rm \Delta\Omega}d\Omega \int\limits_{\rm LOS} ds \frac{\fpbh\,\rho_{\rm DM}(r(s, d, \theta))}{\mpbh} \frac{d^2N_{\gamma}}{dE_{\gamma}dt}\equiv\frac{\fpbh}{4\pi\mpbh}D(\Delta\Omega)\frac{d^2N_{\gamma}}{dE_{\gamma}dt}\\ \nonumber
&  D(\Delta \Omega) \equiv \int\limits_{\Delta \Omega} \int\limits_{\rm LOS} \: \rho_{\rm{DM}}(r(s, d, \theta)) \:ds \: d\Omega \\ \nonumber
& \frac{d^2N_{\gamma}}{dE_{\gamma}dt} = \frac{1}{2\pi}\frac{\Gamma_{\gamma}(E_{\gamma},M_{\rm BH},m)}{e^{E_{\gamma}/T_{\rm BH}}-1} \\ \nonumber
& T_{\rm H} = 1/(4\pi/G_{N\rm }M_{\rm BH})\simeq 1.06\times(10^{16} \rm g / M_{\rm BH})\,\mbox{MeV}
\end{align}

Here $\Gamma_\gamma$ is a grey-body factor and $\Delta\Omega$ is the angular size of the object (or the instrument's FoV, if smaller). The total strength of the signal is determined by the $D(\Delta\Omega)$-factor corresponding to the total mass of the dark matter on the line of sight (LOS) to the object (identical to the $D$-factor considered in the decaying dark matter searches). The spectral shape of the signal is given by $d^2N_{\gamma}/(dE_{\gamma}dt)$-term corresponding to the spectrum peaking at energy $E_\gamma\simeq 5.7\,T_{\rm H}$~\cite{MacGibbon:2007yq} and decreases as a power law for $E_\gamma \ll T_{\rm H}$, see Fig.~\ref{fig:model_signal}. For the results presented below we derived the spectral part of the spectrum with the help of publicly available BlackHawk~\cite{Arbey:2019mbc,Arbey:2021mbl} software for the photons energies between 1 keV and 1 MeV.

We note, that in case of the observations of a distant object the term $D(\Delta\Omega)$ is a growing function of $\Delta\Omega$ that reaches a constant value when the observational region size $\Delta\Omega$ is equal or larger than the characteristic size of a DM halo in the object. Contrary, the contribution from the astrophysical background still increases $\propto \Delta\Omega$ assuming the isotropic background distribution. Thus, the signal-to-noise ratio of the considered signal becomes worse as the size of the observed region exceeds the size of the DM halo of the object. The most efficient observations with a given instrument correspond to the case when the size of the DM halo in the selected object roughly correspond to the instrument's FoV. 

Thus, for the discussed above future instruments, we propose the following targets.
In case of the large-FoV \ep/MXT instrument we propose to observe a MW astrophysical source-free region relatively close to the GC and for $1^\circ$-scale FoV \svom/MXT instrument we propose to focus the observations on the nearby dwarf spheroidal galaxy, e.g. Segue~I (which DM-halo angular size is known to be $\sim 1^\circ$~\cite{Geringer-Sameth:2014yza}).
Due to the relatively complicate instrumental background of both instruments we propose also to perform the observations in ``ON-OFF'' regime, not relying on the background modelling. Namely, we propose that the observation of a dark matter dominated  ``ON'' region (a region relatively close to the GC for \ep/WXT and a Segue~I for \svom/MXT) should be accompanied by the close in time observations of an ``OFF'' region with the lower DM density (a region located far from the GC for \ep/WXT and off-Segue~I observations for \svom/MXT). We require the observations to be performed relatively close in time to minimize the potential impact of the systemactics connected to the time variations of the instrumental background on the derived results. The corresponding  $D$-factors for the proposed observations are indicated in Tab.~\ref{tab:future_missions}.

\begin{figure}
    \centering
    \includegraphics[width=0.45\textwidth]{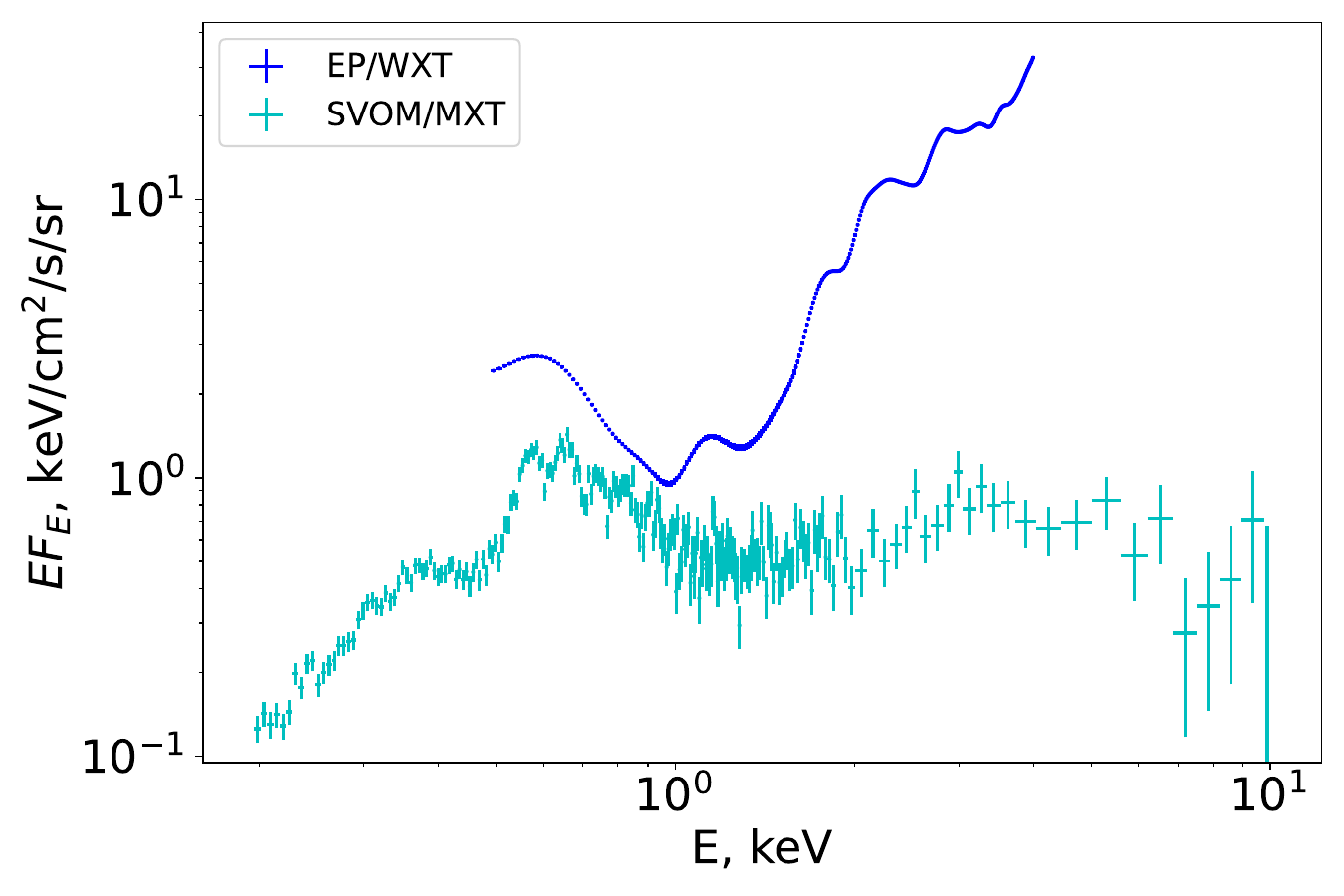}
    \includegraphics[width=0.5\textwidth]{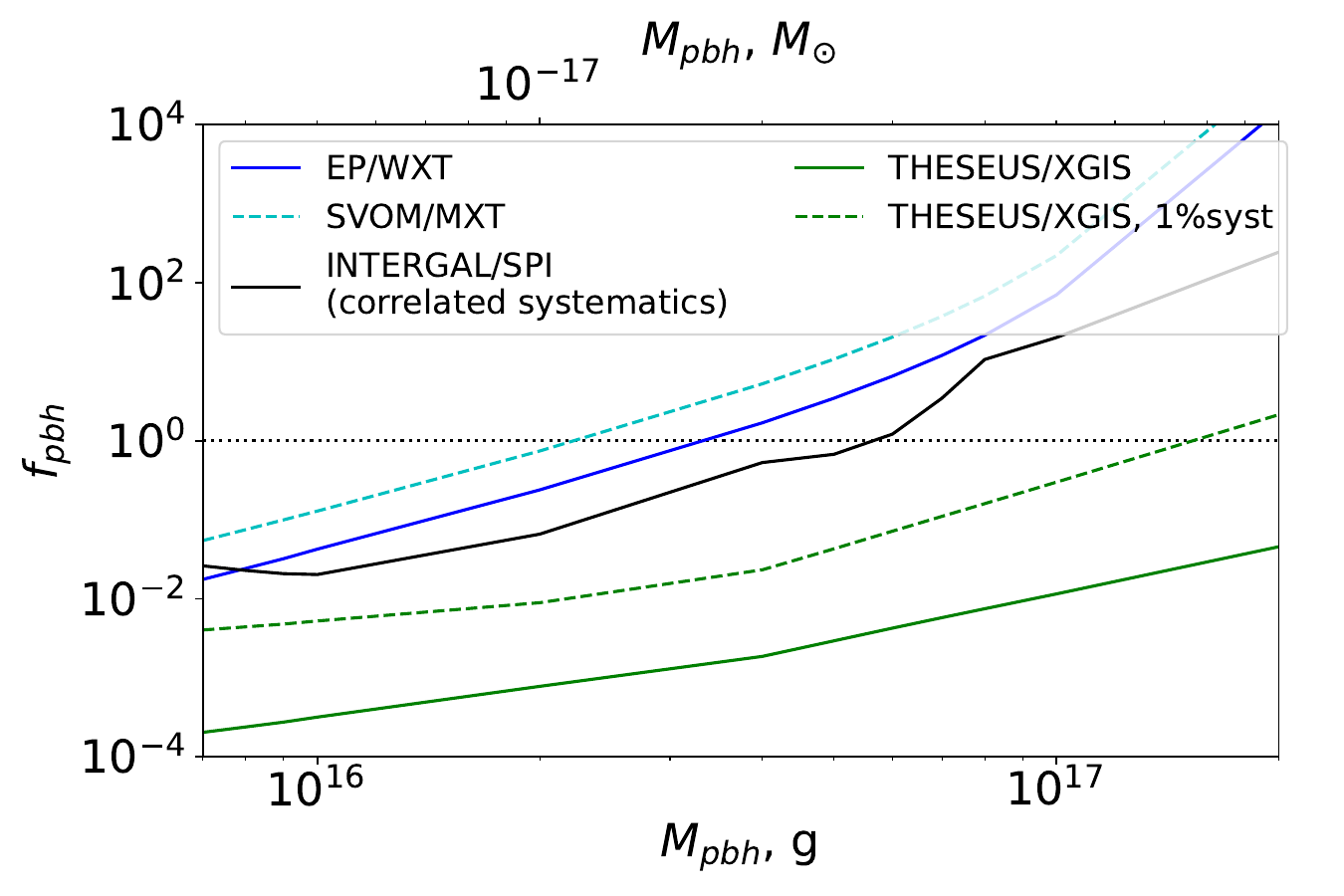}
    \caption{\textit{Left panel:} the spectra ($\propto E^2dN/dE/d\Omega$) of the instrumental+astrophysical backgrounds of the proposed for the observations regions expected to be detected by \ep/WXT (blue) and \svom/MXT (cyan) instruments. \textit{Right panel:} the constraints on evaporating PBHs which could be obtained in case of non-detecting of the Hawking radiation in the keV band with \ep/WXT and \svom/MXT instruments.}
    \label{fig:ep_svom_spi}
\end{figure}

To compare the capabilities of the \ep/WXT and \svom/MXT missions to those discussed in~\cite{pbh_we} we simulated 1~Msec long ON- and OFF-regions observations with these instruments. Both observations included contributions from the astrophysical and instrumental backgrounds for the corresponding instruments\footnote{based on \texttt{epwxt\_bkg.pha} background and \texttt{epwxt.rmf}/\texttt{epwxt\_focus.arf} responses for \ep/WXT and \texttt{MXT\_BG\_20150309\_SandD\_AlOnly\_1150mm.pi} background and \texttt{rmf\_01\_SandD.fits}/\texttt{MXT\_FM\_FULL.arf} responses for \svom/MXT (Dr. D.~Gotz and Dr. Y. Liu private communications) }. The spectra of the simulated OFF-regions were assigned then as backgrounds for the corresponding ON-regions spectra and modelled in XSpec (v.12.13.0c) software. The limits on $\fpbh$ were derived for a set signals corresponding to a set of $\mpbh$ masses, assuming cash-statistics~\cite{cash79} of the residuals (cstat in terms of XSpec). The derived limits correspond to 95\% confidence range statistical limits on $\fpbh$ (corresponding to worsening the fit-statistics by 4.0 ). We note that the described simulation and fitting procedure is similar to the one discussed in~\cite{pbh_we} and allows us the direct comparison of the obtained limits. The ON-regions spectra (the sum of instrumental and astrophysical backgrounds) and the derived limits for \ep/WXT and \svom/MXT instruments are shown in Fig.~\ref{fig:ep_svom_spi}. The right panel illustrates as well the limits based on the INTEGRAL/SPI observations of the regions close to the Galactic Center and expected limits for \ths mission reported in~\cite{pbh_we}.

The right panel of Fig.~\ref{fig:ep_svom_spi} illustrates that \ep/WXT and \svom/MXT instruments with 1~Msec long observations of the proposed regions would be able to provide constraints on evaporating PBHs comparable to those currently derived by the \spi. The best constraints in the discussed PBHs mass range are still expected to be provided by \ths/XGIS mission characterised by significantly higher effective area and broader FoV (in comparison to the discussed here instruments)~\cite{pbh_we}. We note as well that in case of presence of the noticeable systematic uncertainties (connected e.g. to the poor modelling of the time-variable instrumental or a complex astrophysical backgrounds), the constraints from  \ep/WXT and \svom/MXT instruments could worsen and become weaker than the current constraints derived with \spi.

\section{Discussion}

In this proceeding, we discussed the existing limits on the evaporating primordial black holes assuming that they constitute the major fraction of the dark matter. The existing limits based on the evaporation signatures, NS/WD detonation, microlensing events, gravitational-waves and early-universe based constraints effectively cover almost all range of PBH masses see Fig.~\ref{fig:all_constraints}. The only open window in the mass range in which PBHs still can significantly contribute to the dark matter is $3\cdot 10^{16} - 10^{18}$~g. 

We show that the PBHs with masses in this range are expected to evaporate via Hawking radiation producing a signal peaked in keV-MeV band. The intensity of the signal scales with the dark matter amount in the considered object and rapidly decreases with the increase of $\mpbh$, see Fig.~\ref{fig:model_signal}. The limits based on the non-detection of the Hawking radiation signal in keV-MeV band from the GC vicinity and Draco dSph galaxy with \spi and \xmm allowed~\cite{pbh_we} to start to probe the PBH mass range $\lesssim 3\cdot 10^{16}$~g excluding PBHs with such masses as contributors of more than 10\% of the dark matter, see e.g. Fig.~\ref{fig:ath_extp}. 

The next-generation keV-MeV missions that are expected to see the first light within the next decade are expected to be characterised by significantly increased effective area and/or broader FoV. The searches for the signal from the evaporating PBHs with these missions could be potentially interesting for the probing the currently open window in PBHs mass range. 

Following~\cite{pbh_we} we discuss the capabilities of \ath, \extp and \ths missions for such searches, see Tab.~\ref{tab:future_missions} for the basic characteristics of these missions. In addition we study the constraining potential of \ep/WXT and \svom/MXT instruments expected to be launched in 2023-2024. We discussed that the most effective observational strategy for the broad FoV missions (e.g. \extp, \ths and \ep/WXT) would be the ``ON-OFF'' observations of the MW regions with the relatively large amount of the dark matter. The optimal strategy for the narrow-FoV instruments (\ath and \svom/MXT) would be the observation of a certain DM-dominated object with the characteristic size of DM halo close to the size of the FoV of the discussed instruments. 

We show that among the discussed missions the best potential for the improvement of the existing limits on $\fpbh$ and extending the constrained PBH mass range has \ths/XGIS instrument, see Fig.~\ref{fig:ath_extp} and Fig.~\ref{fig:ep_svom_spi} for the expected constraints that could be reached with this instrument and their comparison to the constraints that could be derived with the other discussed instruments. In general, among the discussed missions the tightest limits were obtained for the broad-FoV missions.
At the same time, narrow FoV, excellent energy resolution, and effective area instruments (\ath, \extp/LAD, SFA) are ideal for the search for line-like signals from decaying dark matter~\cite{we_athena,we_extp,we_theseus} are unlikely to provide %compatible
competitive limits in case the systematics for the broad-FoV instruments will be controlled at $\lesssim 1$\% level. The impact of the systematic uncertainty, connected e.g. to the poor modelling of the complex astrophysical and/or time-variable instrumental background could be the key-obstacle for the improving the existing constraints with the broad-FoV missions.

We illustrate the potential reach of the constraints with the next-generation missions (\ths/XGIS) with the green doted-line bordered region labeled ``xgis'' in Fig.~\ref{fig:all_constraints}. Along with the other existing constraints the keV-MeV observations that potentially could be performed within the next decade could allow either detection of evaporating PBHs or will substantially shrink the mass range in which PBHs could constitute the majority of the dark matter.

\section*{Acknowledgements}
The authors acknowledge support by the state of Baden-W\"urttemberg through bwHPC. DM work was supported by DLR through grant 50OR2104 and by DFG through the
grant MA 7807/2-1.

%%%%%%%%%%%%%%%%%%%%%%%%%%%%%%%%%%%%%%%%%%%%%%%%%%%%%%%%%%%%%%%%%%%%%%%%%%%
% Bibliography and bibfile
\def\aj{AJ}%
          % Astronomical Journal
\def\actaa{Acta Astron.}%
          % Acta Astronomica
\def\araa{ARA\&A}%
          % Annual Review of Astron and Astrophys
\def\apj{ApJ}%
          % Astrophysical Journal
\def\apjl{ApJ}%
          % Astrophysical Journal, Letters
\def\apjs{ApJS}%
          % Astrophysical Journal, Supplement
\def\ao{Appl.~Opt.}%
          % Applied Optics
\def\apss{Ap\&SS}%
          % Astrophysics and Space Science
\def\aap{A\&A}%
          % Astronomy and Astrophysics
\def\aapr{A\&A~Rev.}%
          % Astronomy and Astrophysics Reviews
\def\aaps{A\&AS}%
          % Astronomy and Astrophysics, Supplement
\def\azh{AZh}%
          % Astronomicheskii Zhurnal
\def\baas{BAAS}%
          % Bulletin of the AAS
\def\bac{Bull. astr. Inst. Czechosl.}%
          % Bulletin of the Astronomical Institutes of Czechoslovakia
\def\caa{Chinese Astron. Astrophys.}%
          % Chinese Astronomy and Astrophysics
\def\cjaa{Chinese J. Astron. Astrophys.}%
          % Chinese Journal of Astronomy and Astrophysics
\def\icarus{Icarus}%
          % Icarus
\def\jcap{J. Cosmology Astropart. Phys.}%
          % Journal of Cosmology and Astroparticle Physics
\def\jrasc{JRASC}%
          % Journal of the RAS of Canada
\def\mnras{MNRAS}%
          % Monthly Notices of the RAS
\def\memras{MmRAS}%
          % Memoirs of the RAS
\def\na{New A}%
          % New Astronomy
\def\nar{New A Rev.}%
          % New Astronomy Review
\def\pasa{PASA}%
          % Publications of the Astron. Soc. of Australia
\def\pra{Phys.~Rev.~A}%
          % Physical Review A: General Physics
\def\prb{Phys.~Rev.~B}%
          % Physical Review B: Solid State
\def\prc{Phys.~Rev.~C}%
          % Physical Review C
\def\prd{Phys.~Rev.~D}%
          % Physical Review D
\def\pre{Phys.~Rev.~E}%
          % Physical Review E
\def\prl{Phys.~Rev.~Lett.}%
          % Physical Review Letters
\def\pasp{PASP}%
          % Publications of the ASP
\def\pasj{PASJ}%
          % Publications of the ASJ
\def\qjras{QJRAS}%
          % Quarterly Journal of the RAS
\def\rmxaa{Rev. Mexicana Astron. Astrofis.}%
          % Revista Mexicana de Astronomia y Astrofisica
\def\skytel{S\&T}%
          % Sky and Telescope
\def\solphys{Sol.~Phys.}%
          % Solar Physics
\def\sovast{Soviet~Ast.}%
          % Soviet Astronomy
\def\ssr{Space~Sci.~Rev.}%
          % Space Science Reviews
\def\zap{ZAp}%
          % Zeitschrift fuer Astrophysik
\def\nat{Nature}%
          % Nature
\def\iaucirc{IAU~Circ.}%
          % IAU Cirulars
\def\aplett{Astrophys.~Lett.}%
          % Astrophysics Letters
\def\apspr{Astrophys.~Space~Phys.~Res.}%
          % Astrophysics Space Physics Research
\def\bain{Bull.~Astron.~Inst.~Netherlands}%
          % Bulletin Astronomical Institute of the Netherlands
\def\fcp{Fund.~Cosmic~Phys.}%
          % Fundamental Cosmic Physics
\def\gca{Geochim.~Cosmochim.~Acta}%
          % Geochimica Cosmochimica Acta
\def\grl{Geophys.~Res.~Lett.}%
          % Geophysics Research Letters
\def\jcp{J.~Chem.~Phys.}%
          % Journal of Chemical Physics
\def\jgr{J.~Geophys.~Res.}%
          % Journal of Geophysics Research
\def\jqsrt{J.~Quant.~Spec.~Radiat.~Transf.}%
          % Journal of Quantitiative Spectroscopy and Radiative Trasfer
\def\memsai{Mem.~Soc.~Astron.~Italiana}%
          % Mem. Societa Astronomica Italiana
\def\nphysa{Nucl.~Phys.~A}%
          % Nuclear Physics A
\def\physrep{Phys.~Rep.}%
          % Physics Reports
\def\physscr{Phys.~Scr}%
          % Physica Scripta
\def\planss{Planet.~Space~Sci.}%
          % Planetary Space Science
\def\procspie{Proc.~SPIE}%
          % Proceedings of the SPIE
\let\astap=\aap
\let\apjlett=\apjl
\let\apjsupp=\apjs
\let\applopt=\ao
\setlength{\bibsep}{0pt plus 0.3ex} %fix spaces in the bibliography
\bibliography{biblio}

\bigskip
\bigskip
\noindent {\bf DISCUSSION}

\bigskip
\noindent {\bf Dheeraj Pasham:} What exactly are you measuring with \xmm?\\\\
A: To derive the constraints on $\fpbh$ based on \xmm observations of Draco dSph we are measuring the spectrum of diffuse emission in the direction to this dSph. This spectrum is a sum of a several components: astrophysical (hot plasma in the MW; hot plasma in the Solar system); instrumental (power law-like spectrum not convolved with the effective area); PBH dark matter (modelled as discussed above). The observed spectrum is well described with the model including only first two spectral components (astrophysical + instrumental). The non-detection of the dark matter component (with the flux proportional to $\fpbh$) allowed us to put constraints on $\fpbh$.\\
\vskip 1cm
\bigskip
\noindent {\bf John Bally:} Why is there a broken power law in your models on evaporating black holes on the Rayleagh-Jeans tail? I thought Hawking radiation looks like a black body. Are you considering $e^\pm$ pair production above 511~keV or some similar secondary process?\\ \\
A: Yes, correct, the low-energy power-law tail (see Fig.~\ref{fig:model_signal}) originates from the secondary particles emission, mainly $e^\pm\rightarrow\gamma$, see e.g. Fig.4 in~\cite{Arbey:2021mbl}.

\end{document}